\documentstyle[12pt]{article}

\begin{document}
\thispagestyle{empty}

\begin{center}
               RUSSIAN GRAVITATIONAL SOCIETY\\
               INSTITUTE OF METROLOGICAL SERVICE \\
               CENTER OF GRAVITATION AND FUNDAMENTAL METROLOGY\\

\end{center}
\vskip 4ex
\begin{flushright}
                                         RGS-VNIIMS-006/97
                                         \\ physics/9705019

 \end{flushright}
\vskip 15mm

\begin{center}
{\large\bf  Comments on the Hypothesis about Possible Class of Particles
Able to Travel Faster then Light: Some Geometrical Models }

\vskip 5mm
{\bf
M.Yu. Konstantinov }\\
\vskip 5mm
     {\em VNIIMS, 3-1 M. Ulyanovoy Str., Moscow, 117313, Russia}\\
     e-mail: melnikov@fund.phys.msu.su  \\
\end{center}
\vskip 10mm

\begin{abstract}
The hypothesis about possible existence of new class of particles able to
travel faster then light as a source of dark matter, recently formulated by
L. Gonzalez-Mesters, is analyzed. To this end the general geometrical model
for several kinds of matter with different Lorentziann structures coexisting
on the same manifold is introduced and the local energy density in
cosmological reference frame is calculated in two particular cases. It is
shown that the local energy density is positive in both considered cases and
hence such models really may describe cosmological dark matter.
Nevertheless, some problems may appear during the construction of the
cosmological models or the models of the compact objects. Moreover, the
simplest generalization of the model lead to some variants of vector-tensor
theory of gravitation with preferable reference frame which contradict to
observations.
\end{abstract}

\vskip 10mm

\vskip 30mm

\centerline{Moscow 1997}
\pagebreak


\section{Introduction}

In this note we discuss the hypothesis about new class of particles able to
travel faster then light as a possible source of dark matter, which was
proposed recently by L.Gonzalez-Mestres \cite{superlum}. According to this
hypothesis, the apparent Lorentz invariance of classical space-time can be
viewed as a symmetry of the equations of motion of observed matter. In this
case no reference to absolute properties of space and time is required and
besides the usual particles and fields, whose speed could not exceed the
speed of light $c$, an another kinds of matter may exists, whose particles
have limiting speed $c_1>>c$. It was argued, that if there is no direct
interaction between such two (or more) kinds of matter then their
coexistence does not violate the apparent Lorentz invariance of the laws of
physics but may provide most of the cosmic dark matter and produce very high
energy cosmic rays compatible with some unexplained discoveries (see for
example \cite{superlum} and references therein). This hypothesis was
analyzed in \cite{superlum} on pure heuristic level without consideration of
possible mechanisms of gravitational interaction between different sorts of
matter.

In principle, the hypothesis about coexistence of different noninteracting
kinds of matter may be associated with some ideas connecting with formalism
of quantum gravity. Namely, the space-time in classical gravity has metric
with Lorentzian signature while the formalism of continual integration of
quantum gravity uses metric with Euclidean signature. The connection between
Euclidean and Lorentzian metric is realized by Wick rotation. There are
different points of view on this procedure, beginning from the consideration
of Wick rotation as a pure formal method until Hawking's assumption about
Euclidean nature of space-time \cite{hawking78}. If this assumption is
valid, then it is naturally to suppose, that among the knowing kinds of
matter with apparent Lorentzian structure of space-time the other sorts of
matter with their own Lorentzian structures may exist.

In another context the assumption about possible coexistence of different
sorts of matter on the same manifold was formulated in \cite{myuk85} where
the geometrical model for four noninteracting classes of matter with
different Lorentzian structures was given. It was pointed out lately, that
the existence of matter with nonstandard Lorentzian structure may give
contribution into the energy-momentum tensor \cite{myuk94}, but the
corresponding models were not analyzed in details.

In this note the geometric ideas of papers \cite{myuk85,myuk94} are used to
analyze the hypothesis \cite{superlum}. To this end in the next section the
general geometrical background for different Lorentziann structures which
correspond to different sorts of matter on the same Riemannian manifold will
be formulated. The resulting energy-momentum tensor is considered in the
section 3 for two particular models when the light cone of the usual matter
are in the interior of the ''light cone'' of the ''superluminal'' matter and
when the ''light cones'' of usual and ''superluminal'' matter have empty
intersection. Section 4 contains some concluding remarks and discussion.

\section{The multiple Lorentziann structures on the Riemannian manifold}

It is clear, that in the most general form the hypothesis about coexistence
of two or more different Lorentzian structures on the same manifold, which
correspond to the different classes of matter, leads to some kind of bi- or
multi-metric gravitation theory. To provide the gravitational interaction
between particles and fields corresponding to different Lorentziann
structures, the metrics, which define these structures, must be connected
with each other. In this paper the particular form of such connection is
used.

Our approach is based on the well known correspondence between Riemannian
structure of manifold, the field of line elements on it and the Lorentziann
structures on the same manifolds. Namely, let $G_{\alpha \beta }$ is a
Riemannian metric on manifold, $u_\alpha $ is a unit vector field ($%
G^{\alpha \beta }u_\alpha u_\beta =1$) representing the field of line
elements and $c=const>0$ is a light speed, then
\begin{equation}
\label{riem-lormetr}g_{\alpha \beta }=(c^2+1)u_\alpha u_\beta -G_{\alpha
\beta }
\end{equation}
is Lorentziann metric on the same manifold. To apply the above equation to
the analysis of the correspondence between several Lorentziann structures on
the same Riemannian manifold, let's suppose that space-time has Riemannian
metric $G_{\alpha \beta }$ with signature $(+,+,+,+)$ and the apparent
Lorentzian structure, which is associated with observed matter, is defined
by equation (\ref{riem-lormetr}) with the light speed $c=1$ (in geometrical
units). Vector field $u_\alpha $ in (\ref{riem-lormetr}) may be considered
as generator of cosmological reference frame. According to the hypothesis
\cite{superlum} let's suppose also that among the usual Lorentziann
structure there is additional Lorentziann structure on the same manifold,
which correspond to nonobservable (dark) matter and is defined by unit
vector field $v_\beta \neq u_\alpha $ and the ''light speed'' $c_1>1$.

By such a way we obtain a bimetric-type model, whose metrics $g_{\alpha
\beta }$ and $q_{\alpha \beta }$, corresponding to the observed and
hypothetical dark matter, are defined by
\begin{equation}
\label{metr1}g_{\alpha \beta }=2u_\alpha u_\beta -G_{\alpha \beta }
\end{equation}
and
\begin{equation}
\label{metr2}q_{\alpha \beta }=(c_1^2+1)v_\alpha v_\beta -G_{\alpha \beta }
\end{equation}
Following to \cite{superlum}, the particles which moves in Lorentziann
metric $q_{\alpha \beta }$ will be called ''superluminal'' or ''dark
matter''.

It is easy to see that the connection between the determinants of the
metrics $G_{\alpha \beta }$, $g_{\alpha \beta }$ and $q_{\alpha \beta }$ are
given by (see also ):
\begin{equation}
\label{dets}G=-g=-q/c_1^2
\end{equation}
The correspondence between metrics (\ref{metr1}) and (\ref{metr2}) are
defined by\footnote{%
An analogous equations were used also in \cite{beilinson} to generate exact
solutions of vacuum Einstein equations.}
\begin{equation}
\label{corresp}q_{\alpha \beta }=g_{\alpha \beta }-2u_\alpha u_\beta
+(c_1^2+1)v_\alpha v_\beta
\end{equation}

For the following it is necessary to define also the correspondence between
the inverse metrics $g^{\alpha \beta }$ and $q^{\alpha \beta }$. It is clear
that
\begin{equation}
\label{inv1}g^{\alpha \beta }=2u^\alpha u^\beta -G^{\alpha \beta }
\end{equation}
and
\begin{equation}
\label{inv2}q^{\alpha \beta }=\frac{1+c_1^2}{c_1^2}v_G^\alpha v_G^\beta
-G^{\alpha \beta }
\end{equation}
where $u^\alpha =G^{\alpha \beta }u_\beta =g^{\alpha \beta }u_\beta $, $%
v_G^\alpha =G^{\alpha \beta }v_\beta $. Using (\ref{inv1}) equation (\ref
{inv2}) may be rewritten as
\begin{equation}
\label{invcorr}q^{\alpha \beta }=g^{\alpha \beta }-2u^\alpha u^\beta +\frac{%
1+c_1^2}{c_1^2}k^{\alpha \beta }
\end{equation}
where
\begin{equation}
\label{k}k^{\alpha \beta }=4u^\alpha u^\beta (u^\rho v_\rho )^2-2(u^\alpha
v^\beta +v^\alpha u^\beta )u^\rho v_\rho +v^\alpha v^\beta
\end{equation}
with $v^\alpha =g^{\alpha \beta }v_\beta $.

Using the well known equations of the bimetric formalism it is easy to
obtain expressions which connect the covariant derivatives with respect to
metrics $G_{\alpha \beta }$, $g_{\alpha \beta }$ and $q_{\alpha \beta }$.

Equations (\ref{metr1})-(\ref{k}) make possible to analyze the gravitational
interaction between of the usual (observable) matter and nonobservable
(dark) matter.

\section{Energy-momentum tensor of dark matter in the simplest model}

For simplicity in this section we consider two massive scalar fields $%
\varphi $ and $\psi $, one of which ($\varphi $) corresponds to the usual
(observed) matter with metric (\ref{metr1}) and the other field ($\psi $)
corresponds to nonobserved (dark) matter whose own Lorentziann structure are
defined by metric (\ref{metr2}). Simplest action functional for such model
may be written in the form
\begin{equation}
\label{action}S=\int \left( R_g+L_\varphi +c_1L_\psi \right) \sqrt{-g}d^4x
\end{equation}
where $R_g$ is the Ricci scalar of the metric (\ref{metr1}), $L_\varphi $
and $L_\psi $ are the Lagrangians of the fields $\varphi $ and $\psi $
correspondingly:
\begin{equation}
\label{lagrfi}L_\varphi =\frac 12g^{\alpha \beta }\varphi ,_\alpha \varphi
,_\beta +\frac 12m_\varphi ^2\varphi ^2
\end{equation}
and
\begin{equation}
\label{lagrpsi}L_\psi =\frac 12q^{\alpha \beta }\psi ,_\alpha \psi ,_\beta
+\frac 12m_\psi ^2\psi ^2
\end{equation}
where $m_\varphi $ and $m_\psi $ are the masses of the corresponding fields
and the coefficient $c_1$ in (\ref{action}) is present because of (\ref{dets}%
). The geometrical units with $c=\kappa =1$, where $\kappa $ is Einsteinian
gravitational constant, is used here. The absence of mixed $\varphi \psi $
terms in (\ref{action}) provide the Lorentz invariance of the motion
equations of the usual matter ($\varphi $) and nonobservability of the
hypothetical matter ($\psi $).

Substitution of (\ref{invcorr}) into (\ref{lagrpsi}) gives
\begin{equation}
\label{lagrpsi1}L_\psi =\frac 12(g^{\alpha \beta }-2u^\alpha u^\beta +\frac{%
1+c_1^2}{c_1^2}k^{\alpha \beta })\psi ,_\alpha \psi ,_\beta +\frac 12m_\psi
^2\psi ^2
\end{equation}
where $u^\alpha =g^{\alpha \beta }u_\beta $ and $k^{\alpha \beta }$ is
defined by (\ref{k}).

It is clear that variation of (\ref{action}) with respect to $\varphi $ and $%
\psi $ gives the usual Klein-Gordon equations for these fields
\begin{equation}
\label{eqfi}\frac 1{\sqrt{-g}}\frac \partial {\partial x^\alpha }\left(
\sqrt{-g}g^{\alpha \beta }\frac \partial {\partial x^\beta }\right) \varphi
-m_\varphi ^2\varphi =0
\end{equation}
and
\begin{equation}
\label{eqpsi}\frac 1{\sqrt{-q}}\frac \partial {\partial x^\alpha }\left(
\sqrt{-q}q^{\alpha \beta }\frac \partial {\partial x^\beta }\right) \psi
-m_\psi ^2\psi =0
\end{equation}
while variation of (\ref{action}) with respect to $g^{\alpha \beta }$ gives
\begin{equation}
\label{einst}G_{\alpha \beta }=T_{\alpha \beta }+c_1\widetilde{T}_{\alpha
\beta }
\end{equation}
Here $G_{\alpha \beta }$ is Einstein tensor corresponding to the metric $%
g_{\alpha \beta }$, $T_{\alpha \beta }$ is the energy-momentum tensor of
classical matter (field $\varphi $) and the additional term $\widetilde{T}%
_{\alpha \beta }$ is the energy-momentum tensor of the field $\psi $ which
may be associated with hypothetical ''dark'' matter moving in metric $%
q_{\alpha \beta }$. So, the presence of the additional kind of matter with
its own Lorentziann structure does not change the local Lorentz invariance
of the field (and motion) equations for the usual matter but give additional
term in the right-hand side of Einstein equation. In general case this term
has the form%
$$
-\widetilde{T}_{\alpha \beta }=\frac 12\left\{ \psi ,_\alpha \psi ,_\beta
-2\left( u_\alpha \psi ,_\beta +u_\beta \psi _\alpha \right) u^\rho \psi
,_\rho \right\} +
$$
$$
2\frac{c_1^2+1}{c_1^2}\left[ \left( u_\alpha \psi ,_\beta +u_\beta \psi
_\alpha \right) u^\rho \psi ,_\rho \left( u^\rho v_\rho \right) ^2+\left(
u^\rho \psi ,_\rho \right) ^2u^\rho v_\rho \left( u_\alpha v_\beta +u_\beta
v_\alpha \right) \right] -
$$
$$
\frac{c_1^2+1}{c_1^2}\left[ \left( u_\alpha \psi ,_\beta +u_\beta \psi
_\alpha \right) v^\rho \psi ,_\rho u^\sigma v_\sigma +\left( v_\alpha \psi
,_\beta +v_\beta \psi _\alpha \right) u^\rho \psi ,_\rho u^\sigma v_\sigma
\right] -
$$
$$
\frac{c_1^2+1}{c_1^2}\left[ u^\rho \psi ,_\rho v^\sigma \psi ,_\sigma \left(
u_\alpha v_\beta +u_\beta v_\alpha \right) -\left( v_\alpha \psi ,_\beta
+v_\beta \psi ,_\alpha \right) v^\sigma \psi ,_\sigma \right] -
$$
\begin{equation}
\label{darktensor}\frac 14g_{\alpha \beta }\left( q^{\alpha \beta }\psi
,_\alpha \psi ,_\alpha +m_\psi ^2\psi ^2\right)
\end{equation}

The above expression is rather complicated. By this reason it will be
considered in two particular cases when the light cons of the usual and
''superluminal'' matter are collinear ( $u_\alpha =v_\alpha $ with $c_1>>1$)
and orthogonal $u_\alpha \perp v_\alpha $ ($c_1=1$) to each other.

{\it 1. Colinear light cones. }In this case, which is closely connected with
the ideas of paper \cite{superlum}, $u_\alpha =v_\alpha $, $c_1>>1$, tensor $%
k^{\alpha \beta }=u^\alpha u\beta $ and (\ref{lagrpsi1}) takes the form
\begin{equation}
\label{case1}L_\psi =\frac 12(g^{\alpha \beta }-\frac{c_1^2-1}{c_1^2}%
u^\alpha u^\beta )\psi ,_\alpha \psi ,_\beta +\frac 12m_\psi ^2\psi ^2
\end{equation}
For the energy-momentum tensor of this field we have the following
expression
\begin{equation}
\label{emt1}-\widetilde{T}_{\alpha \beta }=\frac 12\left\{ \psi ,_\alpha
\psi ,_\beta -\frac{c_1^2-1}{c_1^2}\left( u_\alpha \psi ,_\beta +u_\beta
\psi ,_\alpha \right) u^\rho \psi ,_\rho \right\} -\frac 12g_{\alpha \beta
}L_\psi
\end{equation}
The local energy density in the reference frame which is defined by vector
field $u_\alpha $ is equal to
\begin{equation}
\label{dens1}\widetilde{T}_{\alpha \beta }u^\alpha u^\beta =\left( u^\rho
\psi ,_\rho \right) ^2\frac{c_1^2-2}{2c_1^2}+\frac 14m_\psi ^2\left(
c_1^2+\psi ^2\right)
\end{equation}
Here we use well known equation $q^{\alpha \beta }\psi ,_\alpha \psi ,_\beta
=m_\psi ^2c_1^2$. It is clear, that the local energy density in this case is
positive if $c_1^2>2$, non-negative if $c_1^2=2$ and may be both positive
and negative if $1<c_1^2<2$.

{\it 2. Orthogonal light cones.} In this case $c_1=c=1$, $u^\alpha v_\alpha
=0$, $k^{\alpha \beta }=v^\alpha v^\beta $ and (\ref{lagrpsi1}) reads
\begin{equation}
\label{case2}L_\psi =\frac 12\left( g^{\alpha \beta }-2u^\alpha u^\beta
+2v^\alpha v^\beta \right) \psi ,_\alpha \psi ,_\beta +\frac 12m_\psi ^2\psi
^2
\end{equation}
Direct calculation gives the following expression for energy momentum tensor
of the field $\psi $ in this case
$$
-\widetilde{T}_{\alpha \beta }=\frac 12\left\{ \psi ,_\alpha \psi ,_\beta
-2\left( u_\alpha \psi ,_\beta +u_\beta \psi ,_\alpha \right) u^\rho \psi
,_\rho \right\} +
$$
\begin{equation}
\label{emt2}\left( v_\alpha \psi ,_\beta +v_\beta \psi ,_\alpha \right)
v^\rho \psi ,_\rho -\frac 12g_{\alpha \beta }L_\psi
\end{equation}
The local energy density in the reference frame which is defined by vector
field $u_\alpha $ (in cosmological reference frame) is equal to
\begin{equation}
\label{dens2}\widetilde{T}_{\alpha \beta }u^\alpha u^\beta =\frac 32\left(
u^\rho \psi ,_\rho \right) ^2+\frac 14m_\psi ^2\left( 1+\psi ^2\right)
\end{equation}
It is easy to see, that the local energy density in this case is positive
because $u^\rho \psi ,_\rho $ and $m_\psi ^2$ can not become zero
simultaneously.

The last model may be generalized as follows \cite{myuk85}. Let vector field
$u_\alpha $ generate Lorentzian structure which corresponds to the observed
matter and let the fields $v_\alpha ^i$ ($i=1,2,3$) form an orthonormal
frame in orthogonal space sections, i.e. $G^{\alpha \beta }v_\alpha
^iv_\beta ^j=\delta ^{ij}$ and $u^\alpha v_\alpha ^i=0$. Analogously to the
above, it may be supposed that every field $v_\alpha ^i$ corresponds to some
kinds of non-observable (''dark'') matter with its own Lorentziann structure%
$$
q_{\alpha \beta }^i=2v_\alpha ^iv_\beta ^j-G_{\alpha \beta }=g_{\alpha \beta
}-2u_\alpha u_\beta +2v_\alpha ^iv_\beta ^j
$$
In this case the full energy-momentum tensor $T_{\alpha \beta }$ will
consist of four terms: the term corresponding to the usual matter and three
additional terms which correspond to the nonobserved matter. In the simplest
case when hypothetical kinds of matter are represented by massive scalar
fields these additional terms will have the form (\ref{emt2}).

\section{Concluding remarks}

In the modern classical general relativity metric of space-time depends on
the matter distribution. However the signature of the metric, which defines
the local Lorentziann structure of space-time, is supposed to be fixed. It
may be supposed that the local Lorentziann structure of space-time is also
defined by the features of matter. If such supposition is true then the
question about uniqueness of such structure is naturally appears. This
question may be reformulated also as the question about possible coexistence
on the same manifold several different classes of matter which generate
different non-equivalent Lorentziann structures.

As it was argued in \cite{superlum}, the coexisting of such different
classes of matter without direct interaction between them does not
contradict to the apparent Lorentz invariance of the laws of physics. On the
other hand it may provide most of the cosmic dark matter and produce very
high energy cosmic rays compatible with some unexplained discoveries (see
for example \cite{superlum} and references therein).

It is shown that such different classes of matter, which is mentioned in\cite
{superlum}, may be naturally described using the well known correspondence (%
\ref{riem-lormetr}) between Riemannian structure of manifold (Riemannian
metric $G_{\alpha \beta }$), the field of line elements on it (represented
by the unit vector field $u_\alpha $) and the Lorentziann structures
(Loretzian metric $g_{\alpha \beta }$) on the same manifold. Using this
correspondence, two Lorentzian structures $g_{\alpha \beta }$ and $q_{\alpha
\beta }$ on the same manifold $(M^4,G)$ may be connected by equation (\ref
{corresp}), which may be used for description of the gravitational
interaction between the usual matter, which moves in metric $g_{\alpha \beta
}$, and hypothetical matter moving in metric $q_{\alpha \beta }$. Using the
simplest model it was shown that the presence of the additional kind of
matter with its own Lorentziann structure does not change the local Lorentz
invariance of the field (and motion) equations for the usual matter but
gives additional term in the right-hand side of Einstein equation. This
additional term may be considered as the energy-momentum tensor of dark
matter.

In the case then hypothetical dark matter is represented by massive scalar
field $\psi $ its energy-momentum tensor is given be equation (\ref{emt1})
for colinear light cones or (\ref{emt2}) for orthogonal light cones. The
corresponding local energy density in cosmological reference frame connected
with vector field $u_\alpha $ is given by equations (\ref{dens1}) and (\ref
{dens2}) respectively. In both cases the local energy density is positive,
so both models may describe cosmological dark matter. It is clear, that in
the case of colinear light cones the homogeneous isotropic cosmological
solutions for metric $g_{\alpha \beta }$ may be easily constructed as well
as in the standard general relativity while in the case of orthogonal light
cones metric $g_{\alpha \beta }$ may be spatially homogeneous in this case
only if the metrics $q_{\alpha \beta }^i$ are static. On the other hand,
there are no obvious problems with construction of the models of the compact
objects for the usual matter with the metric $g_{\alpha \beta }$ in the case
with orthogonal light cones, while the case of colinear light cones may lead
to the apparent violation of the equivalence principle because any
concentration of the usual (observable) matter is an attractor for dark
matter and vice versa.

The action (\ref{action}), which where considered in the previous section,
is non-symmetric with respect to metrics $g_{\alpha \beta }$ and $q_{\alpha
\beta }$: action (\ref{action}) may be varied with respect to metric $%
g_{\alpha \beta }$ or with respect to metric $q_{\alpha \beta }$ but not
both. Moreover, the fields $u^\alpha $ and $v^\alpha $ remain undefined and
their nature is unclear. The introduction of these fields in Lagrangian as
usual massive or massless vector fields leads to some variant of
vector-tensor gravitation theory with preferable reference frame, whereas it
is known that the existence of such frames contradicts to
observations~\cite{will}.

Thus, the hypothesis \cite{superlum} that the cosmological dark matter is
generated by the hypothetical particles which able to travel faster then
light has rather simple geometrical realization, but its consistency with
observations is problematic.

\section*{Acknowledgments}

This work was supported by the Russian Ministry of Science and the Russian
Fund of Basic Sciences (grant N 95-02-05785-a).

\end{document}